\documentstyle[12pt]{article}

\begin{document}

\newcommand{\beq}{\begin{equation}}
\newcommand{\eeq}{\end{equation}}

\newcommand{\beqn}{\begin{eqnarray}}
\newcommand{\eeqn}{\end{eqnarray}}

\newcommand{\su}{$ SU(2) \times U(1)\,$}

\newcommand{\np}{Nucl.\,Phys.\,}
\newcommand{\pl}{Phys.\,Lett.\,}
\newcommand{\pr}{Phys.\,Rev.\,}
\newcommand{\prl}{Phys.\,Rev.\,Lett.\,}
\newcommand{\prep}{Phys.\,Rep.\,}
\newcommand{\zp}{Z.\,Phys.\,}

\newcommand{\eps}{\epsilon}
\newcommand{\mw}{M_{W}}
\newcommand{\mww}{M_{W}^{2}}
\newcommand{\mz}{M_{Z}}
\newcommand{\mzz}{M_{Z}^{2}}

\newcommand{\epm}{$e^{+} e^{-}\;$}

\newcommand{\ra}{\rightarrow}
\newcommand{\lra}{\leftrightarrow}
\newcommand{\tr}{{\rm Tr}}

\newcommand{\gag}{$\gamma \gamma$ }

\newcommand{\dkg}{\Delta \kappa_{\gamma}}
\newcommand{\dkz}{\Delta \kappa_{Z}}
\newcommand{\dz}{\delta_{Z}}
\newcommand{\dgz}{\Delta g^{1}_{Z}}
\newcommand{\la}{\lambda}
\newcommand{\lag}{\lambda_{\gamma}}
\newcommand{\laz}{\lambda_{Z}}
\newcommand{\lnl}{L_{9L}}
\newcommand{\lnr}{L_{9R}}
\newcommand{\lt}{L_{10}}
\newcommand{\lu}{L_{1}}
\newcommand{\ld}{L_{2}}

\newcommand{\cl}{{{\cal L}}}
\newcommand{\cd}{{{\cal D}}}
\newcommand{\cv}{{{\cal V}}}
\def\slashc{c\kern -.400em {/}}
\def\slashL{L\kern -.450em {/}}
\def\slashcl{\cl\kern -.600em {/}}
\def\W{{\bf W}}
\def\B{{\bf B}}
\def\noi{\noindent}
\def\sm{${\cal{S}} {\cal{M}}\;$}
\def\nph{${\cal{N}} {\cal{P}}\;$}
\def\ssb{${\cal{S}} {\cal{S}}  {\cal{B}}\;$}
\def\cviol{${\cal{C}}\;$}
\def\pviol{${\cal{P}}\;$}
\def\cpviol{${\cal{C}} {\cal{P}}\;$}

\begin{titlepage}
\rightline{August 1993}
\rightline{ENSLAPP-A-431/93}
\rightline{hep-ph/9308343}

\begin{center}

\vskip 1.cm

{ \Large \bf
Tests on the Hierarchy of Trilinear and Quadrilinear Weak Bosons
Couplings at the NLC
and  Comparison with the LHC/SSC
\footnote{Talk given at the ``Workshop on Physics and Experiments with
Linear $e^+e^-$ Collider", Waikoloa, Hawaii, April 26-30, 1993. This is
a longer version of the contribution to the proceedings.\\}}

\vspace*{2cm}

{\bf Fawzi Boudjema }
          \vspace{0.5cm} \\

{\it Laboratoire de Physique Th\'eorique}
EN{\large S}{\Large L}{\large A}PP
\footnote{URA 14-36 du CNRS, associ\'ee \`a l'E.N.S de Lyon, et au L.A.P.P.
d'Annecy-le-Vieux}\\
{\it Chemin de Bellevue, B.P. 110, F-74941 Annecy-le-Vieux, Cedex, France.}

\end{center}

\vspace*{2cm}

\begin{abstract}

I first review a few basic guiding principles that lead to the notion
of a hierarchy of couplings in searches of New Physics involving weak
bosons processes. The hierarchies within a linear and a non-linear realization
of symmetry breaking are compared to the usual {\em phenomenological}
parameterization of the $WWV$ vertex. Limits that one expects to obtain
at the NLC(500GeV) and LHC/SSC on the tri-linear and quadri-linear
anomalous $W$
couplings are compared. The cleanness of an  \epm gives the
NLC a clear advantage in constraining the tri-linear couplings. However, with
``only" $500$GeV, the \epm is not competitive with $pp$ colliders in
probing the quadri-linear couplings. An interpretation in terms of scalar-like
and vector-like models is given and I argue that in the presence of a
light (or ``not-so-heavy") Higgs, the New Physics affecting the $W$ sector
would be easier to pin-down with a moderate energy \epm
machine.

\end{abstract}

\end{titlepage}

\newpage

\section{W Physics and the Higgs Connection}

\baselineskip=14pt

The $W$ system as described by the \sm (Standard Model)
reflects the marriage of two
fundamental principles:\\
\noi $\star \star $ Gauge Principle in its non-Abelian form \\
\noi $\star \star $ Spontaneous Symmetry Breaking which provides
the $W$ and $Z$ bosons with mass.\\
\noi These two items taken together are a quite unusual combination
as you may convince yourself if you go through the Particle Data Book:
the $W$ would constitute the {\em only} known system of massive and
gauge spin-1 particles. Yet,
it must be stressed that to date there
has been {\em no} \underline{direct} tests of either the local non-Abelian
nature nor of the exact realization of \ssb (Spontaneous Symmetry Breaking)
in the W system. \\
\noi
The gauge principle leads most straightforwardly to the universality of the
weak coupling
constant, that is, all couplings involving the $W\&Z$ are the same allowing
for the quantum numbers. The non-Abelian local gauge symmetry tells us that
the couplings of the W to fermions is the same as the tri-linear W coupling
as well as the quadri-linear coupling. The couplings of the $W$ and $Z$ to
fermions have been
tested to a high degree of accuracy directly. However, these tests can be, in a
sense, regarded as direct {\em Abelian} tests.
Indirect limits on the 3-W and the
4-W self-couplings have been worked out through their effects on quantum
corrections, but these limits are plagued with ``theoretical error bars",
interpretations and ambiguities. \\
\noi
The tri-linear as well as the quadrilinear couplings of the $W$ come solely
 from the (generalized) $SU(2)$ kinetic term through the field strength,
$\W_{\mu \nu}$
\beqn
\W_{\mu \nu}=\frac{1}{2} \left(
\partial_\mu \W_{\nu}- \partial_\nu \W_{\mu} +\frac{i}{2} g [\W_\mu, \W_\nu]
\right)=\frac{\tau^i}{2} \left(\partial_\mu W^{i}_\nu-\partial_\nu W^{i}_\mu
-g \epsilon^{ijk}W_{\mu}^{j}W_{\nu}^{k} \right)
\eeqn
\noi
(with $\W_\mu=W^{i}_{\mu}\tau^i$, the normalization for the Pauli
matrices is $\tr(\tau^i \tau^j)=2\delta^{i j}$).
The Abelian hypercharge field does not generate any tri-linear couplings.
Defining
\beqn
\B_{\mu \nu}=\frac{1}{2} \left(
\partial_\mu B_{\nu}- \partial_\nu B_{\mu} \right)\tau_3
\;\;\;\; \B_\mu=\tau_3 B_\mu
\eeqn
\noi the kinetic term writes
\beqn
\cl_{{\rm Gauge}}=- \frac{1}{2} \left[
\tr(\W_{\mu \nu} \W^{\mu \nu}) + \tr(\B_{\mu \nu} \B^{\mu \nu}) \right]
\eeqn
\noi So far, in a sense, this only describes the {\em transverse} $W$'s.
Save for the quantum numbers and group assignments the construction
is as the one used for $QCD$.
The longitudinal $W$'s, as well as the mixing between the left and right
fermionic states
are to be revealed in the ${\em mass}$ terms.

\subsection{Longitudinal $W$'s and the inclusion of mass}
\noi
The reason that the longitudinal degrees of freedom do not
effecaciously contribute to
the above Lagrangian can be gleaned by recalling that a longitudinal
state of polarization for the $Z$, say, of momentum $k$ may be written as
\beqn
\epsilon_{\mu}^{L}
=\frac{k_\mu}{M_Z} - M_Z \frac{s_\mu}{s.k} \;\;{\rm with} \; \;
s^2=0
\eeqn
which exhibits the all-important high-energy leading behaviour
($k_\mu/M_Z \sim E_Z/M_Z)$. This also shows
that a longitudinal $Z$ could be represented as the gradient of a scalar field
$Z^{L}_{\mu} \propto \partial_\mu \phi_3$. It is clear that $Z^L$ written
this way, $Z^L(\phi_3)$,
does not contribute to the kinetic term since $Z^{L}_{\mu \nu}(\phi_3)=
\partial_\mu Z^L_\nu - \partial_\nu Z^L_\mu=0$.
However, it contributes to the mass.
The mass terms of concern to us here are
\beqn
\cl_M=M_W^2 W^+_\mu  W^{-\mu} + \frac{1}{2}M_Z^2 Z_\mu Z^\mu
\eeqn
\noi Put by hand, on its own, this term breaks the gauge invariance. Rather,
it completely {\em hides} it.
To introduce the longitudinal modes in a manifestly gauge invariant way,
one exploits the fact that the longitudinal mode may be regarded as the
gradient of a scalar field. One then has to turn this gradient into a
{\em \bf covariant derivative}.

\subsection{Symmetry Breaking: The SM option}
\noi
For the $W$'s one needs three of these (pseudo)-scalars.
One then has to ``group" them,
{\em i.e.}, find a representation for them. Here, we are helped by another
very well
confirmed experimental measurement. The $\rho$ parameter is to the per-mil
level equal to 1.
\noi
This means that in the absence of mixing with the hypercharge,
the $W^\pm$ and $W^0$ have the same mass. This corresponds to an extra global
$O(3) \approx SU(2)$ symmetry, termed $SU(2)_c$ custodial symmetry, which
manifests itself
in the scalar sector. It so happens that the most simple representation of
the scalars has this symmetry. In the \sm one introduces a complex doublet,
$\Phi$, with hypercharge $Y=1$,
\beqn
\Phi=\frac{1}{\sqrt{2}} \left( \begin{array}{c}
\phi^+\\
\phi_0\end{array} \right)\equiv exp(\frac{i \omega^i \tau^i}{v})
\left( \begin{array}{c} 0 \\ \frac{v+H(x)}{\sqrt{2}} \end{array}
\right)\;\; {\rm and}\;\;
\cd_\mu \Phi=\left(\partial_\mu +\frac{i}{2} (g \W_\mu +g'Y B_\mu) \right)\Phi
\eeqn
$\omega_i$ are the Goldstone Bosons,
$\cd_\mu \Phi $ is the covariant derivative on $\Phi$.
This gives the most general {\em renormalizable} Lagrangian
\beqn
\cl_{H,M}&=&(\cd_\mu \Phi)^\dagger(\cd_\mu \Phi)\;-\;
\lambda \left[ \Phi^\dagger \Phi - \frac{\mu^2}{2 \lambda}
\right]^2
\eeqn
As is well known, when one goes to the unitary gauge not only do we recover the
above mass terms but also the interaction of the Higgs scalars. Therefore the
study of the interaction in the $W$ sector is a window on the mechanism of
symmetry breaking.
%
%

\subsection{Symmetry Breaking: The Non-Linear
Realization \protect\cite{AppelquistLong}}
One can also uncover the gauge invariance of the mass terms even in the
eventuality that the Higgs does not exist.
Instead of using the dim-1 field $\Phi$, one appeals to the (dim-0)
matrix $\Sigma$
which only describes the Goldstone Bosons with the built-in custodial $SU(2)_c$
symmetry:
\beqn
\Sigma=exp(\frac{i \omega^i \tau^i}{v}) \;\; (v=246 \;GeV {\rm \;is \;
the \;vev)} \;\;{\rm and}\;\;
{{\cal D}}_{\mu} \Sigma=\partial_\mu \Sigma + \frac{i}{2}
\left( g \W_{\mu} \Sigma
- g'B_\mu \Sigma \tau_3 \right)
\eeqn
The gauge invariant form of the mass terms is made explicit by the use of
the {\em covariant} derivative and the $\Sigma$-``field" through the operator
of order ${{\cal O}}(p^2)$ (2-because it involves two derivatives)
\beqn
\cl_M=\frac{v^2}{4} \tr(\cd^\mu \Sigma^\dagger \cd_\mu \Sigma)
\equiv - \frac{v^2}{4}  \tr\left( \cv_\mu \cv^\mu \right) \;\;\
{\rm with}\;\;
\cv_\mu=\left( {{\cal D}}_{\mu} \Sigma \right) \Sigma^{\dagger}
\eeqn
The unitary gauge is obtained by formally setting $\Sigma \ra${\bf 1}.
Of course, with the non-linear realization one ends up with a
{\em non-renormalizable} model. At one-loop the ensuing divergences
are only logarithmic and can be associated with the Higgs mass dependence
of the \sm low-energy observables.
This construction is important because it shows that seemingly non-invariant
operators can be made gauge invariant without recourse to the Higgs particle, a
point which has been stressed some time ago
\cite{AppelquistLong} and has been revived recently \cite{Cliff}. For
the record I would like to quote a sentence from a lecture given by Appelquist
13 years ago \cite{Appelquist}:\\
\begin{center}``{\em The massive Yang Mills theory is formally equivalent to
the non-linear Lagrangian ......with the advantage of being straighforward to
analyze dimensionally and being easily regularized by the linear model}".
\end{center}
%
%
This important reminder should still be kept in mind when trying to criticize
phenomenological parameterization of New Physics, \nph, in the bosonic sector.
The use of the covariant derivative or the field strength,
which is nothing else
but the commutator of two covariant derivatives, will
give a gauge invariant description, even with operators
beyond the \sm . We only have to decide about
the Higgs content in order to choose between a linear or a non-linear
realization of \ssb. What should also transpire from these considerations,
is that the probing of the self-interactions of the $W$ and the search for
any departure from the minimal structure is a test of the symmetry breaking
especially if the Higgs persists to be elusive.

\section{Anomalous Weak Bosons Self-Couplings: Para-
meterizations and Classifications}

\subsection{The standard ``phenomenological" parameterization of the tri-linear
coupling}
One knows \cite{Majorana}
that a particle of spin-J which is not its own anti-particle
can have, at most, $(6J+1)$
electromagnetic form-factors including \cviol, \pviol and \cpviol
violating terms. The
same argument tells us \cite{Majorana}
that if the ``scalar"-part of a massive spin-1
particle does not contribute, as is the case for the Z in
$e^+ e^- \ra W^+ W^-$,
then there is also the same number of invariant form-factors for
the spin-1 coupling to a charged spin-J particle. This means that
there are 7 independent $WWZ$ form factors and 6 independent $WW\gamma$
form-factors beside the electric charge of the W.  This number of invariants
is derived by appealing to angular momentum conservation and to the
conservation of the {\em Abelian} $U(1)$ current: {\em i.e.}, {\bf two} utterly
established symmetry principles one would, at no cost, dare to tamper with.
Although one can not be more general than this, if all these $13$
couplings were simultaneously allowed on the same footing, in an experimental
fitting procedure and most critically to the best probe $e^+e^- \ra W^+W^-$,
it will be
a formidable task to disentangle between all the effects, or to extract
good limits on all. \\
\noi One then asks
whether other symmetries, though not as inviolable as the two previous ones,
may be invoked to reduce
the set of permitted extra parameters. One expects that the more contrived
a symmetry has thus far been verified, the less likely a parameter which breaks
this symmetry is to occur, compared to a parameter which respects these
symmetries.
For instance, in view
of the null results on the electric dipole moments of fermions and other
\cpviol violating observables pointing to almost no \cpviol violation,
\cpviol violating terms, and especially the electromagnetic ones, are very
unlikely to have any detectable
impact on W-pair production.
Therefore, in a first analysis they should not
be fitted. The same goes for the \cviol violating $WW\gamma$ couplings.
Additional symmetry principles and then theoretical ``plausibilty
arguments" can be invoked to further reduce the parameter space of the
anomalous
couplings. However, before invoking any additional criteria other than
angular momentum conservation, conservation of the {\em Abelian} current,
unobservable \cpviol and electromagnetic \cviol violation, we should give the
{\em \bf most} general {\em phenomenological} parameterization of the
$WWV$ vertex. This parameterization is to be used at {\em tree-level} in
processes describing vector boson pair production by light fermions
(or any other crossed channels of these). It assumes the vector bosons to be
either {\em on-shell} or associated to a conserved current.
With this warning....

\subsubsection{\cviol and \pviol conserving $WWV$ couplings}

\noi There are now two parameterizations on the market and some confusion
between the defining parameters has, unfortunately, arisen, especially as
concerns the parameter $\kappa_Z$. Below, I give the two
parameterizations and the conversion between the two.
The oft-used parameterization of Hagiwara et al. \cite{HPZH},
(the {\em HPZH} parameterization) is
\beqn
{\cal L}_1= &-ie& \left\{ \left[ A_\mu \left( W^{-\mu \nu} W^{+}_{\nu} -
W^{+\mu \nu} W^{-}_{\nu} \right) +
\overbrace{ (1+\mbox{\boldmath $\Delta \kappa_\gamma$} )}^{\kappa_\gamma}
F_{\mu \nu} W^{+\mu} W^{-\nu} \right] \right.
\nonumber \\
&+& \left. cotg \theta_w \left[\overbrace{(1+ {\bf \Delta g_1^Z})}^{g_1^Z}
Z_\mu \left( W^{-\mu \nu} W^{+}_{\nu} -
W^{+\mu \nu} W^{-}_{\nu} \right) +
\overbrace{(1+\mbox{\boldmath $\Delta \kappa_Z$} )}^{\kappa_Z}
Z_{\mu \nu} W^{+\mu} W^{-\nu} \right] \right.
\nonumber \\
&+& \left. \frac{1}{M_{W}^{2}}
\left( \mbox{\boldmath $\lambda_\gamma$} \;F^{\nu \lambda}+
\mbox{\boldmath $\lambda_Z$} \;
cotg \theta_w Z^{\nu \lambda}
\right) W^{+}_{\lambda \mu} W^{-\mu}_{\;\;\;\;\;\nu} \right\}
\eeqn
The {\em BMT} Collaboration \cite{BMT}
has preferred the use of the following couplings
\beqn
{\cal L}_1= &-ie& \left\{ \left[ A_\mu \left( W^{-\mu \nu} W^{+}_{\nu} -
W^{+\mu \nu} W^{-}_{\nu} \right) +
\overbrace{ (1+\mbox{\boldmath $x_\gamma$} )}^{\kappa_\gamma}
F_{\mu \nu} W^{+\mu} W^{-\nu} \right] \right.
\nonumber \\
&+& \left. \overbrace{(cotg \theta_w +\mbox{\boldmath $\delta_Z$}
)}^{{\bf g_{WWZ}}}
\left[ Z_\mu \left( W^{-\mu \nu} W^{+}_{\nu} -
W^{+\mu \nu} W^{-}_{\nu} \right) +
\overbrace{(1+ \bf{ \frac{x_Z}{g_{WWZ}} } )}^{\kappa'_Z}
Z_{\mu \nu} W^{+\mu} W^{-\nu} \right] \right.
\nonumber \\
&+& \left. \frac{1}{M_{W}^{2}}
\left( \mbox{\boldmath $y_\gamma$} \;F^{\nu \lambda} + \mbox{\boldmath $y_Z$}
\;
Z^{\nu \lambda}
\right) W^{+}_{\lambda \mu} W^{-\mu}_{\;\;\;\;\;\nu} \right\}
\eeqn
The conversion is given by
\beqn \label{conversion}
x_\gamma=\Delta \kappa_\gamma \;\;;\;\;\
\delta_Z=\frac{c_w}{s_w}\Delta g^1_Z \;\;;\;\;
x_Z=\frac{c_w}{s_w}(\Delta \kappa_Z -\Delta g^1_z) \;\;;\;\;
y_\gamma=\lambda_\gamma \;\;;\;\;y_Z=\frac{c_w}{s_w} \lambda_Z
\eeqn

\noi Coming back to the warning
about the use of this phenomenological parameterization outside its context,
for instance to vector boson scattering. Even at tree-level it
should be modified/extended to include appropriate accompanying
``anomalous" quartic couplings. This is especially acute
for $\lambda$ and $g_1^Z$, to restore
$U(1)_{em}$ gauge invariance, at least...\\

\subsubsection{\cpviol preserving but \pviol violating operators}

\noi The inclusion of the other operators assumes violation of \cviol and/or
\pviol. These may be searched for only if one reaches excellent
statistics.  Therefore the next
operator which may be added is the \cpviol conserving but
\pviol-violating Z coupling.
In the {\em HPZH} parameterization \cite{HPZH}
this coupling is introduced through $g_5^Z$
\beqn
\cl_2=- e (\frac{c_w}{s_w} g_5^Z) \; \epsilon^{\mu \nu \rho \sigma}
\left(W_\mu^+ (\partial_\rho W_\nu) - (\partial_\rho W_{\mu}^{+}) W_\nu \right)
 Z_\sigma
\eeqn
\subsection{A natural hierarchy of couplings through gauge invariance and
scaling}

\noi Recently this general parameterization has been fiercely attacked on the
ground that it does not respect the full local \su gauge invariance
\cite{Ruj}.
By now, recalling the introductory remarks, the parry to the criticism should
be immediate. It is the same
as for the mass term, i.e, \underline{2W} coupling:
the general Lagrangian written above is but a
{\em particular}
parameterization written in a {\em specific} gauge where only the
{\em physical}
fields are kept and only those parts describing {\em tri-linear couplings} are
exhibited \cite{Espriu}.
As what has been done for the mass term we can always rewrite any
of the above parameters within a gauge invariant operator
\cite{Cliff,Fernand}. This is achieved by
extensively using the {\em covariant} derivative and specifying how to
represent the Goldstone Bosons. For the latter
specification one would, essentially, be making an assumption about
the ``lightness" of the Higgs.
Unless, of course, the Higgs has already been discovered. At the $500$GeV NLC
one will not wait too long to know....\\

\noi The important point about a gauge-invariant
formulation is that it greatly extends the
domain of application of the anomalous parameters, ....even at the
quantum level.
\noi However, imposition of local symmetries
alone is not sufficient to reduce the number of parameters.
What has been making the success of present-day physical
theories, which apart from masses have only a few parameters, is
not just their built-in  gauge invariance. It is also because
they are
caracterized by the {\em lowest} dimension of all possible gauge invariant
operators.
This makes them {\em renormalizable} and predictive. Higher dimension
operators destroy the power of unequivocal predictive calculability as one
needs more and more inputs from experiments. However, the very fact that these
higher dimension operators are, necessarily, inversely proportional
to the scale ($\Lambda$) of \nph, that they parameterize,
means that
their effect at low energy is small. They contribute with a penalising
factor ($(E/\Lambda)^{n-4}$), where $E$ is the typical low energy of the
particular process and $n$ is the dimension of the operator.
Therefore, due to the limited accuracy in our experiments one can only hope
to see the effect of the next dimension operators which will be referred to as
next-to-leading or sub-leading operators. The leading being, of course,
those of the \sm (with or without the Higgs).
Higher
order, or sub-sub-leading operators, (with even larger $n$)
are even less likely to have any impact.
There
is a tacit assumption here, namely that the coefficients of the
operators are not too large so that an expansion in energy is possible.
\noi
This is the scaling argument augmented in the case of spin-1's with the gauge
principle. This is really ``Wilsonian" in spirit \cite{Wilson}:
\begin{center} {\em ``The couplings should have an order of importance, and
for any desired but given degree of accuracy only a finite subset of the
couplings would be needed".}\end{center}

The  most straighforward illustration of these notions is provided by
a very simple example. This is the Lagrangian describing photons at energies
much below the electron mass. It is also interesting because in a sense
it describes anomalous self-couplings of the photon (bilinear and quartic).

\subsubsection{Interlude: The Effective Lagrangian for Photons Below the
``Electron Threshold" }

\noi Imagine a world with just {\em massless}
photons and that we want to write the most
general Lagrangian with the only information or rather stricture
being the local $U(1)$ gauge invariance.
If we require the Lagrangian
to be renormalizable then the only operator possible is the kinetic term
below, it is the {\em marginal} operator
\footnote{In the sense of being equally important at all energies.}
. The only
problem with this example is that this term
does not represent  any interaction, it is a free-field trivial theory.
Interactions between photons is possible through the introduction
of higher order effective operators.
These would be the impact the ``heavy"
unobservable electrons will leave at these lilluputian energies. Because
the symmetry we have at these energies is the $U(1)$ local symmetry the only
possibility to describe any of these interactions is to use the
electromagntic field strength (or its dual). These are dimension-2 objects
and scale as the energy (or frequency) of the photon. The first terms in
(in principle infinite) set of operators is \cite{Euler}
\beqn
\cl_{eff.}^{QED}&=& -\frac{1}{4} F_{\mu \nu}F^{\mu \nu}
+ \frac{\beta_1}{m^2} \frac{e^2}{16 \pi^2} \left( F_{\mu \nu} \Box F^{\mu \nu}
+\frac{\epsilon_2}{m^2} F_{\mu \nu} \Box^2 F^{\mu \nu} \right) \nonumber \\
&+& \frac{1}{m^4} \frac{e^4}{16 \pi^2} \left(
\beta_2 (F_{\mu \nu} F^{\mu \nu})^2
\;+\;\beta_3 (F_{\mu \nu} \tilde{F}^{\mu \nu})^2 \right) + ......
+\cl_{gauge\;fixing}
\eeqn
(Note that we have added a gauge-fixing term so that we can invert the
photon propagator.)\\
\noi
The first anomalous operator, characterized by $\beta_1$,
is a correction to the
two-point function. If one were to make an analogy with $W$ physics
this kind of self-energy operators can be extremely contrived
 from LEP1 measurements. I have not included any tri-linear anomalous
couplings.
This is not forbidden by gauge invariance, in fact I have taken a theoretical
biais: for the three-neutral particles one needs to break \cviol invariance
which
is not possible also in the fundamental theory: QED with electrons only.
The first genuine interaction is a quartic coupling which describes
the scattering of light by light.\\
These operators scale as the inverse of the ``scale of New Physics". In this
particular case this must be related to the electron mass.
At energies much lower
than this scale the effect of these operators is small since the presence of
higher and higher derivatives means that the corrections are of order
$(\omega/m_e)^n$ for a dim-n operator, where $\omega$ is a typical
photon frequency. Therefore only very few of
the next to lowest operators are necessary to have a good enough precision.
Since we know that QED is the fundamental theory which gives rise to this
effective Lagrangian we have definite predictions for the values of all
the $\beta_i$. These low-energy values are
obtained by taking the full QED lagrangian and considering one-loop
diagrams. One then expands the results in the limit of a very large
electron mass, $m_e$. It should be kept in mind that in the process,
a renormalization procedure has been carried out, which
among other things {\em defines} and specifies the value of $\alpha$. \\
\noi With $m=m_e$ one finds a specific pattern between the ``anomalous
couplings" emerging from QED:
\beq
\beta_1=6 \beta_2=\frac{24}{7} \beta_3=\frac{1}{15} \;\;\; ; \;\;\;
\epsilon_2=\frac{3}{28}
\eeq
For further reference one should note that in this particular example
the heavy particle (or \nph) has been ``integrated out" at one -loop. One can
think of other effective field theories where the non-renormalizable
terms result from integrating out
a heavy particle at tree-level, the Fermi four-point interaction is one
notable example. This is the reason I have choosen to pull out, from
the definition of the $\beta_i$'s, factors of $1/16\pi^2$ which
betray their (one-) loop origin. The factors of $e^2$
comes from the observation that each photon field contributes a factor of
$e$ and is a reflection of the fact that in the fundamental theory charged
particles couple to the photon with the ``universal" strength $e$. For
effective operators not describing gauge particles universal
coupling factors are not contained in the coefficients. Moreover, for
operators
describing heavy particle tree-level exchanges, there is no reason to include
the factor $1/16\pi^2$ with the expectation that these operators have a more
significant impact than those corresponding to loop effects. \\
The term $\epsilon_2$
can be considered as the first order term in the expansion of
the ``form factor", $\beta_1$.

{\em \noi The Question of Loops}

\noi This is an aside.
One might wonder whether one should use the higher dimension operators
inside loop diagrams. The worry is that since these are
high-derivative operators, power counting indicates that they
have a high degree of divergence such that the positive power in
the cut-off (proportional to the scale) introduced to regularize these
diagrams ``overcomes" the (inverse) power of the mass which scales
the operators. One might conclude that these operators do not decouple.
This would be
very unnatural, in fact, as shown and argued by many \cite{Cliff,Einhorn},
any such divergence can be absorbed in the {\em definition} of
the parameters of the Lagrangian.
For the example at hand, $\beta_2$ contributes
to the $q^2$ part of the self-energy and $\beta_2$ is used to define a
running of $\alpha$. The quartic couplings $\beta_{2,3}$ could through
one loop (by joining two of the photon lines) be turned into a two-point
function. It is easy to see that the induced vertex (a tadpole-type)
is quartically divergent
and contributes to the kinetic term. This divergence can be easily disposed
off by redefining (rescaling)
the electromagnetic field. The regularization procedure
in this very simple example does not turn out to be so crucial. In more complex
situations it is more practical to use a regularization which respects the
symmeties
of the Lagrangian, otherwise the regularization procedure can introduce
spurious
terms which destroy the original symmetries. In any case, what is more
important is to have all the symmetries
and the ensuing Ward (BRST) identities. We could then use any regularization.
The
application of the Ward identities will show which of the divergences
are an artifact
of the regularization. These terms can then be removed by the
introduction of (additional) counterterms.

\subsubsection{Back to the W system:
The Ranking of the Gauge Invariant Operators}
\noi  There are two important concepts in the ranking
of the operators in the two approaches of \ssb. In the linear approach the
classification is done according to the dimension of the operator, i.e,
to the
power of the scale, $\Lambda$, of the \nph. In the non-linear scenario
this is done on the basis of a
momentum expansion. Therefore, for the next-to-leading (or most ``probable")
operators, of order
${{\cal O}}(p^4)$, a new scale does not necessarily appear. \\
\noi One expects the scale of \nph which ``weighs" the anomalous operators
to be larger than $\sim TeV$. Therefore the ``sub-sub-leading"
operators should not be considered.
If the NLC(500) is to run after the LHC(SSC) one would, by then, know whether
this is correct....
\noi There is another symmetry to be included when listing the most
likely operators: the custodial $SU(2)_c$ global symmetry.

\noi On the basis of the above symmetries, one can not help it, but
there are operators which contribute to
the tri-linear couplings and have a part which corresponds to
bi-linear anomalous
$W$ self-couplings. Because of the latter and of the unsurpassed precision
of LEP1, these operators are already very much {\em unambiguously}
constrained. I will
not list them. I will only list the ones which we have not had direct
access to as they have no bi-linear part. These are the operators which
in the parlance
 of \cite{Ruj} are referred to as ``blind directions".
I do this with a pervading feeling of uneasiness since one must admit that it
is very hard to come up with theories which only give rise to the latter
or where the former are very much suppressed.
With these few points spelled out, we arrive at the most {\em probable}
set of yet-untested operators, within a linear \cite{BuchWy,Ruj} or a
non-linear \cite{AppelquistLong,Holdom,Espriu,FLS,BDV,Feruglio,AppelquistWu}
realization of \ssb.
\newpage
\begin{table}[h]
\caption{\label{linearvsnonlinear}
{\em The Next-to-leading Operators describing the $W$ Self-Interactions which
do not contribute to the $2$-point function.}}
\vspace*{0.3cm}
\centering
\begin{tabular}{|l||l|}
\hline
{\bf Linear Realization \hspace*{0.1cm}, \hspace*{0.1cm} Light Higgs}&
{\bf Non Linear-Realization \hspace*{0.1cm}, \hspace*{0.1cm} No Higgs}\\
\hline
&\\
${{\cal L}}_{B}=i g' \frac{\epsilon_B}{\Lambda^2} (\cd_{\mu}
\Phi)^{\dagger} B^{\mu \nu} \cd_{\nu} \Phi$&
${{\cal L}}_{9R}=-i g' \frac{L_{9R}}{16 \pi^2} \tr ( {\bf B}^{\mu \nu}\cd_{\mu}
\Sigma^{\dagger} \cd_{\nu} \Sigma )$ \\
&\\
${{\cal L}}_{W}=i g \frac{\epsilon_w}{\Lambda^2} (\cd_{\mu}
\Phi)^{\dagger} (2 \times \W^{\mu \nu}) (\cd_{\nu} \Phi)$&
${{\cal L}}_{9L}=-i g \frac{L_{9L}}{16 \pi^2} \tr ( \W^{\mu \nu}\cd_{\mu}
\Sigma \cd_{\nu} \Sigma^{\dagger} ) $ \\
&\\
$\cl_{\lambda} = \frac{2 i}{3} \frac{L_\lambda}{ \Lambda^2}
g^3 \tr ( \W_{\mu \nu} \W^{\nu \rho} \W^{\mu}_{\;\;\rho})$&
$\;\;\;\;\;\;\;\;---------\;\;\;\;$\\
&\\
$\;\;\;\;\;\;\;\;---------\;\;\;\;$&
$\cl_{1}=\frac{L_1}{16 \pi^2} \left( \tr (D^\mu \Sigma^\dagger D_\mu \Sigma)
\right)^2\equiv \frac{L_1}{16 \pi^2} {{\cal O}}_1$ \\
$\;\;\;\;\;\;\;\;---------\;\;\;\;$&$
\cl_{2}=\frac{L_2}{16 \pi^2} \left( \tr (D^\mu \Sigma^\dagger D_\nu \Sigma)
\right)^2 \equiv \frac{L_2}{16 \pi^2} {{\cal O}}_2$ \\
& \\
\hline
\end{tabular}
\end{table}

\noi We see that the combination of gauge invariance, $SU(2)_c$ global
symmetry (only broken by mixing)
and the principle of ``minimality" keeping the leading terms in the
energy expansion does not give any \cviol or \pviol violation. Excluding
electromagnetic \cpviol violation, $SU(2)_c$ suffices to forbid
\cpviol violation also for the $Z$. This is a strong argument for assuming
$SU(2)_c$.
Moreover, in the
non-linear realization the counterpart of $L_\lambda$ is relegated to a lower
cast as it is counted as ${{\cal O}}(p^6)$:$L_\lambda \propto
\tr \left( \left[{{\cal D}}_{\mu},{{\cal D}}_{\nu} \right]
\left[{{\cal D}}^{\nu},{{\cal D}}^{\rho} \right]
\left[{{\cal D}}_{\rho},{{\cal D}}^{\mu} \right] \right)$.
This operator as we will see contributes to $\lambda$ in the phenomenological
parameterization. Another view is that this operator is not really telling us
much about symmetry breaking. It involves in a sense only transverse $W$'s.
If there were no \ssb, i.e, if the $W$ had no mass, $L_\la$ would be the only
operator that we would write. Although, for the tri-linear couplings,
there are more paramaters in the linear
realization, one can, in principle, perform more tests (in Higgs production)
with $\cl_{B,W}$ than with $\cl_{9R,9L}$ as anomalous Higgs-W vertices are
also induced by $\cl_{B,W}$.
On the other hand, the operators $L_{1,2}$ which represent genuine
quartic couplings (they do not contribute to the tri-linear couplings)
and involve a maximum number of longitudinal modes are
sub-sub-dominant in the light Higgs scenario. When relinquishing the
Higgs, $L_{1,2}$ would be the most important manifestation of alternative
symmetry breaking scenarios. Unfortunately,
they can not be
probed in $e^+e^- \ra W^+W^-$. Note that $L_{9L,W}$ contributes a
part to the $4V$ vertex. We will come back to these quartic couplings
later.\\

\subsubsection{The most likely $WWV$ couplings}

\noi By going to the
physical gauge, one recovers the phenomenological parameters with the
{\em constraints}:
\beqn \label{constraints}
\kappa_{\gamma}-1&=&\Delta\kappa_\gamma=x_\gamma=
\frac{e^2}{s_w^2} \frac{v^2}{4 \Lambda^2}
(\epsilon_W+\epsilon_B )=
\frac{e^2}{s_w^2} \frac{1}{32 \pi^2} \left( L_{9L}+L_{9R} \right)
\nonumber \\
\kappa_{Z}-1&=&\Delta\kappa_Z=\frac{e^2}{s_w^2} \frac{v^2}{4 \Lambda^2}
(\epsilon_W -\frac{s_w^2}{c_w^2} \epsilon_B)=
\frac{e^2}{s_w^2} \frac{1}{32 \pi^2} \left( L_{9L}
-\frac{s_w^2}{c_w^2} L_{9R} \right)
\nonumber \\
g_{1}^{Z}-1&=&\Delta g_1^Z=\frac{e^2}{s_w^2} \frac{v^2}{4 \Lambda^2}
(\frac{\epsilon_W}{c_w^2})=
\frac{e^2}{s_w^2} \frac{1}{32 \pi^2} \left(\frac{ L_{9L}}{c_w^2} \right)
\nonumber \\
\lambda_\gamma&=&\lambda_Z=\left(\frac{e^2}{s_w^2}\right)
L_\lambda \frac{M_W^2}{\Lambda^2}
\eeqn
Note that these couplings, in the {\em BMT} \cite{BMT} notation, verify the
custodial global symmetry.
$x_Z c_w =- x_\gamma s_w $. Only $\cl_{W,9L}$ give $\delta_Z$ ($\Delta g_1^Z)$
with $\delta_Z=\frac{x_\gamma}{s_w c_w}$. In the numerical applications I will
take $\alpha$ and ``$s_w$" at $M_Z^2$, i.e, in Equation ~(\ref{constraints})
$e \ra e(M_Z^2)$ and
$s_w^2 \ra s_Z^2=0.228$. Not that there is a one-to-one correspondence
$L_{9L,9R}\leftrightarrow \eps_{W,B}$ for the $WWV$ parts. So, for two bosons
production or neglecting Higgs exchanges in $3V$ production, the two sets are
equivalent (same constraints).
%
%
\\

\noi
\subsubsection{Next-to-best: Breaking the Global Symmetry and
Maintaining the ``Order"}

\noi The order, here, is the order in the energy expansion or the
dimensionality
of the operators. To make the point, I stick with the non-linear realization.
We know that there is a slight breaking of the global $SU(2)_c$, for instance,
as
induced by the top. The first effect appears in the $2W$ vertex and contributes
to the $\rho$ parameter. Introducing $X=\Sigma \tau^3 \Sigma^\dagger$,
the contribution to $\Delta \rho$ is through the leading ${{\cal O}}(p^2)$
operator
\beqn
\cl_{\Delta \rho}= \Delta \rho \frac{v^2}{8}
\left( \tr ({{\cal V}}_\mu X ) \right)^2
\eeqn
A typical $3W$ $SU(2)$-breaking operator is \cite{Feruglio}
\beqn
\slashcl_1 =i g \frac{\slashL_1}{16 \pi^2}
\left( \tr ( \W^{\mu \nu} X ) \right)
\left( \tr (X [\cv_\mu,\cv_\nu] ) \right)
\eeqn
and leads to
\beqn
\Delta g_1^Z= \delta_Z=0 \;\;;\;\;
\Delta\kappa_\gamma=x_\gamma=
\frac{e^2}{s_w^2} \frac{1}{32 \pi^2} \left( 4 \slashL_1 \right) \;\;;\;\;
\Delta \kappa_{Z}=
\frac{e^2}{s_w^2} \frac{1}{32 \pi^2} \left( 4 \slashL_1 \right)
\eeqn
\noi so that $\frac{c_w}{s_w} x_\gamma=x_z
\not= -\frac{s_w}{c_w} x_\gamma$. The inequality is a reflection of the
explicit $SU(2)_c$ breaking (in the language of {\em BMT}).
A $WWZ$ \cviol violating but \cpviol conserving
operator at the same order in the energy expansion is also possible when
the custodial symmetry is broken, as first noticed by Feruglio
\cite{Feruglio}. With $\widetilde{\W}^{\mu \nu}=\frac{1}{2}
\eps^{\mu \nu \alpha \beta} \;\W_{\alpha \beta}$
\beqn
\slashcl_\slashc = g \frac{\slashL_\slashc }{16 \pi^2}
\left( \tr ( \widetilde{\W}^{\mu \nu} \cv_\mu ) \right)
\left( \tr (X \cv_\nu ) \right) \ra
g_5^Z= \frac{e^2}{s_w^2} \frac{1}{32 \pi^2} \left(
\frac{ - \slashL_\slashc}{c_w^2} \right) {\rm \;\; with \;\;} g_5^{\gamma}=0
\eeqn

I will, in the remainder, \underline{assume}
that the amount of explicit $SU(2)_c$ breaking beyond that of the \sm is small
so that these effects are of second importance. If not, we see that one has,
with these two additional $SU(2)_c$ breaking operators, the \underline{same}
number
of parameters as with the {\em phenomenological} parameterization
of the \cpviol conserving ``dim-4" $WWV$ vertex. \\

\subsubsection{Quartic Couplings}

\noi Writing the contribution of the ``genuinely quartic" anomalous operators
in the physical gauge, we obtain:
\beqn
\cl_{Q}^{(4)}= \left( \frac{e^2}{s_w^2} \right)^2 \frac{1}{16 \pi ^2}
\left\{ L_1 \left( W^{+\mu} W^{-}_{\mu} W^{+ \nu}W^{-}_{\nu}
+\frac{1}{c_w^2} W^{+}_{\mu}W^{-\mu} Z_\nu Z^\nu
+\frac{1}{4 c_w^4}Z_\mu Z^\mu Z_\nu Z^\nu \right) + \right. \nonumber \\
\left. L_2 \left( \frac{1}{2} (W^{+\mu} W^{-}_{\mu} W^{+ \nu}W^-_{\nu}
+ W^{+}_{\mu} W^{+\mu} W^{- \nu}W^-_{\nu})
+\frac{1}{c_w^2} W^{+}_{\mu}W^{-}_{\nu} Z^\mu Z^\nu
+\frac{1}{4 c_w^4}Z_\mu Z^\mu Z_\nu Z^\nu \right) \right\}
\eeqn
\noi We \cite{BB4V}
have already obtained this form by only appealing to $SU(2)_c$ global.
To make contact with that analysis, the correspondence is
$g_{0,c}=\frac{e^2}{16 \pi^2} \frac{1}{s_w^2} L_{1,2}$.
Neither the
$WWWW$ nor the $WWZZZ$ have a form like that found in the \sm.
But most importantly there is a $ZZZZ$ coupling which is not  present in
the \sm at {\em tree-level}. Also, note that with these genuine
quartic couplings, photons do not appear. The first operator parameterizes the
exchange of a heavy scalar. This point is also crucial, because while
tri-linear couplings could be the residual effect of integrating out heavy
particles at one-loop, the quartic couplings can correspond to integrating
heavy states at tree-level and therefore one would expect their coefficients
to be larger. We also note that in the combination
${{\cal O}}_1-{{\cal O}}_2$, i.e., $L_1=-L_2$,
the $4Z$ vanishes. This corresponds to ``vectorial"
theories, {\em i.e.}, integrating out heavy spin-one. This could be
of relevance to
technicolour models. The quartic part of the $\cl_{9L}$ does induce new
$WWZ\gamma$, $WWZZ$ and $WWWW$ but no $4Z$ ensues:

\beqn
\cl_{9L}^{(4)}
&\ra& \overbrace{ \frac{e^2}{s_w^2}\frac{L_{9L}}
{32 \pi^2} }^{ \Delta \kappa_{\gamma}} \frac{2 e^2}{s_w^2}
\left\{ \frac{s_w}{ c_w} \left(
A_\mu Z^\mu W^{+}_{\nu}W^{-\nu} - \frac{1}{2}
A^\mu Z^\nu ( W^{+}_{\mu}W^{-}_{\nu} + W^{+}_{\nu}W^{-}_{\mu} ) \right)
\right. \\
&+& \left. \left(
Z_\mu Z^\mu W^{+}_{\nu}W^{-\nu} - Z^\mu Z^\nu W^{+}_{\mu}W^{-}_{\nu} \right)
+
\frac{1}{2} \left(
W^{+\mu} W^{-}_{\mu} W^{+\nu} W^{-}_{\nu} -
W^{+ \mu} W^{+}_{\mu} W^{-\nu}W^{-}_{\nu} \right) \right\} \nonumber
\eeqn

\section{Direct Searches at the Next Colliders}

\subsection{$e^+e^- \ra W^+ W^-$}
In \epm colliders the most promising channel to probe the tri-linear
couplings is $W$ pair production, due to the large statistics that it offers.
Here, I will refer to the excellent extensive study conducted in Europe by
the {\em BM2} \cite{BMT} Collaboration and will translate their results within
the effective Lagrangian approach. {\em BM2}
can fit many parameters at a time beside
giving limits
on individual parameters of the {\em phenomenological} Lagrangian assuming
various relations between them. Misha Bilenky has kindly rerun their
program for me within the $SU(2)_c$ symmetric chiral Lagrangian approach, i.e.
by considering the effects of $L_{9L,9R}$. With the conversion in
(~\ref{conversion}) this is the
same constraint as with the parameters $L_{W,B}$. {\em BM2}
take advantage of the
cleanness of the \epm environment to reconstruct a large number of the
$W^+W^-$ density matrix elements (DME). No beam polarization
is exploited in this analysis. They, conservatively, only take into account the
semi-leptonic decays where the leptons are either $e^\pm$ or
$\mu^\pm$ (no $\tau$'s) to have an excellent reconstruction of the
the scattering angle ($\theta$). Apart from the total differential
cross-section, and disregarding any \cpviol violation,
the fits are done on 4 independent combinations of the
DME which do not rely on any charge identification of the final fermions
and 3 additional DME based on the i.d. of the lepton charge
only. This gives them {\bf 8} independent
observables instead of using the full five-fold differential cross-section
for the 4 final fermions.
 Simulated data are generated according to the {\em tree-level}
\sm expectations with an angular coverage of $|\cos \theta|<0.98$, taking
8 bins in this variable while in the variables of the fermions
6 bins are taken, requiring a minimum of 4 events in each bin.
Idealistically though,
only statistical errors are taken into account. The extracted limits are at
the $95\%$C.L. When referring to Fig.1, note that LEP2 represents
$\sqrt{s}=190GeV$ with
$\int \cl =500pb^{-1}$, while the NLC500 is with $\sqrt{s}=500GeV$
and $\int \cl =10fb^{-1}$. We should still keep in mind that, apart from
$RC$, the effect of beamstrahlung has not been included yet.

\subsection{$e^+e^- \ra W^+ W^-\gamma, W^+ W^-Z$}
Triple vector boson production offers the possibility to check for quartic
couplings. Of course, the tri-linear couplings also enter in $WW\gamma$ and
$WWZ$ but {\em NOT} in $ZZZ$ production as would  genuine $SU(2)_c$ symmetric
quartic couplings. The only exception is the $L_V$ ($L_1=-L_2$) realization
of the quartic coupling, but then this would contribute to $WWZ$ only.
Note that all next-to-leading quartic couplings never contribute
to $WW\gamma$. Therefore, by looking in all 3V production channels one can
easily discriminate between genuine quartic and tri-linear couplings.
The table below illustrate this discrimination (the number of stars indicates
the sensisitivity of the particular channel to the operators).Besides, in
case of a signal, the origin of any of the couplings can be unravelled
through characteristic energy and angular
distributions. This is discussed in \cite{BB4V}. Let me point out also, that if
a
tri-linear coupling is to give a detectable signal in $WW\gamma, WWZ$ then
it would give a more prominent effect in $WW$ production, so that in this case
the value of the coupling would be extracted from $W$ pair analysis. It would
then be
included in $WWZ$ production so that one checks whether there is any additional
contribution in this channel.

\begin{table}[t]
\caption{\label{3vvs4v}
{\em Contributions of the Next-to-Leading Operators of the Chiral Lagrangian
to three-vector productions in \epm}}
\vspace*{0.3cm}
\begin{tabular}{|c|c|c|c|c|}
\cline{2-5}
\multicolumn{1}{c|}{}&$e^+e^- \ra W^+W^-$&$e^+e^- \ra W^+W^-\gamma$&
$e^+e^- \ra W^+W^-Z$&$e^+e^- \ra ZZZ$
\\ \hline
$\lnl, L_{W\phi}$&$\star \star \star$&$\star$&$\star$& {\bf NO}\\ \hline
$\lnr, L_{B\phi}$&$\star \star \star$&$\star$&$\star$& {\bf NO}\\ \hline
$L_{\la}$&$\star \star \star$&$\star$&$\star$& {\bf NO}\\ \hline
\hline
$\lu$&{\bf NO}&{\bf NO}&$\star$&$\star$ \\ \hline
$\ld$&{\bf NO}&{\bf NO}&$\star$&$\star$ \\ \hline
$L_V (\lu=-\ld)$&{\bf NO}&{\bf NO}&$\star$&{\bf NO} \\ \hline \hline
\end{tabular}
\end{table}

The interesting aspect about triple vector
production, especially $WWZ$ and $ZZZ$, at a moderate CM energy
is that it is a substitute to $W$ fusion processes which are ineffective
at $500$GeV.
The 3-V cross-sections are not very large though.
For instance, at $\sqrt{s}=500$GeV we
have $\sigma(WWZ)\sim 39$fb. The $ZZZ$ is tiny $\sigma(ZZZ) \sim 1fb$, but this
very fact classifies this reaction as a {\em rare}
process and hence it is a good testing ground for \nph.
Our \cite{BB4V} analysis on
$WW\gamma$ production was done with the following cuts,
$|\eta_\gamma|<2$, $p_T^\gamma >20$GeV, $|\cos (\angle eW)|< 0.96$,
$\cos (\angle \gamma W), \cos (\angle W W)  < 0.985$. With these cuts
the cross-section amounts to $\sim 112$fb. The branching ratio into
$\tau$'s was not considered in any of the $3V$ production. The limits we
extract are based on detecting a $3\sigma$ deviation
only in the total cross-section (including branching fractions). Due to the low
statistics we did not aim at reconstructing the final polarizations and
only statiscal errors were taken into account. For $WWZ$ (and $ZZZ$ !)
$3\nu$ final states were not counted. For $ZZZ$ the discovery criterion was
an excess of signal events as large as that corresponding to  a $3\sigma$
in $WWZ$ since the bulk of the events looks like in $WWZ$ and as
we do not expect invariant mass reconstructions ($M_W$ {\em v.s} $M_Z$) to be
discriminating.
For $WW\gamma$ we have considered the effect of $L_{9L,9R} \equiv L_{W,B}
\equiv (\delta_z, x_\gamma; x_Z=-s_w/c_w x_Z)$ and also
$L_\la (\la_\gamma=\la_Z)$. In $WWZ$ we considered, in addition
to the above couplings, the important effect of the novel
$L_1, L_2, L_V$. Finally in $ZZZ$ only $L_1, L_2$
take part.
For all $3V$ productions we have only taken one parameter at a time and assumed
$\int\cl=10fb^{-1}$.

\subsection{$\gamma \gamma  \ra W^+ W^-$}
If the \epm linear collider is turned into a high-energy/high-luminosity
$\gamma \gamma$ collider through Compton back-scattered laser light, this
process will constitute the largest cross-section. At an effective
$\sqrt{s_{\gamma \gamma}} \sim 500$GeV, $\sigma(WW) \sim 80$pb! The
$\gamma \gamma$ mode would be
a $W$ factory. Of course, one can exploit this to check for anomalous
$WW\gamma$ (and also $WW\gamma \gamma$) couplings without making any assumption
about the $Z$ counterparts. Keeping in line
with my assumptions I will only consider
the case $L_{9L,9R}$ (no $\lambda$ fitting). We know that, in effect, we are
measuring $\Delta \kappa_{\gamma} \propto L_{9L}+L_{9R} \propto L_W+L_B$, so
this is
basically a one parameter fit: the combination $L_{9L}+L_{9R}$. I have
reinterpreted the results obtained by Choi and Schrempp \cite{Choi},
where a realistic
$\gamma \gamma$ luminosity spectrum was considered but without beam
polarization effect. The $WW$ reconstruction efficiency including
branching ratios is taken at $15\%$ based on events with $|cos \theta|<0.7$.
Statistical errors are taken into account and the systematic are estimated.
The bounds are at the $90\%$ CL.

\begin{figure}[t]
\caption{{\em Comparison between the expected bounds on the two-parameter
space $(L_{9L},L_{9R}) \equiv (L_W,L_B) \equiv (\Delta g_1^Z,
\Delta \kappa_\gamma \;\; x_z c_w=-x_\gamma s_w)$ (see text for
the conversions) at the NLC500, SSC/LHC
and LEP2. The NLC bounds are from $e^+e^- \ra W^+W^-\;,W^+W^-\gamma, W^+W^-Z$
(for the latter these are one-parameter fits) and $\gamma \gamma \ra W^+W^-$.
The SSC/LHC bounds are from $pp \ra WZ, W\gamma$. We also show (``bars")
the limits on one single parameter.}}
\vspace*{17cm}
\end{figure}

\clearpage

\subsection{$pp \ra W\gamma$ and $pp \ra WZ$}
For the comparison with the $pp$ machines, I  refer to the analysis conducted
recently with the constrained set $L_{9L,9R}$ \cite{FLS}.
The $q \bar{q} \ra W^+ W^-$
is either fraught with huge QCD backgrounds or in case of the ``all-leptonic"
decay will be very difficult to reconstruct, it will thus offer very
little chance for the study of \ssb. The authors \cite{FLS}
consider $WZ$ and $W\gamma$
production. $WZ$ is a much better channel: both final bosons can be
longitudinal and hence the largest deviations are expected here. The authors
also find that $q \bar q' \ra WZ$ is more efficient than the
$WZ$ fusion process as far as $L_{9L,9R}$ are concerned.
The maximum deviation coming essentially from $W_L Z_L$, it is clear that
$L_{9L}$ is overwhelmingly dominating through its $g_1^Z$ term. The
$\kappa_{\gamma,Z}$ only lead to $W_L Z_T$. In $W\gamma$ once again one probes
$\kappa_\gamma$, that is, the combination $(L_{9L}+L_{9R})$. The samples only
contain the decays into $e\&\mu$. For the $WZ$ channel, bounds are set
by requiring a doubling of events (with at least an excess of
$40$ at the SSC and $30$ at the LHC) in the high-$p_T^Z$ range
$300<p_T^Z<750$GeV. Almost the same criterion is used for $W\gamma$
(but with $400<p_T^\gamma<750$GeV). For both SSC and LHC $\int \cl=10fb^{-1}$.

\subsection{$W$ fusion processes at $pp$ and the quartic couplings}
The best channel to look for the effect of the genuine
quartic couplings is the like-sign $W$ pair production: $W^\pm W^\pm$.
Within the parameterization in terms of $L_{1,2,V}$ a very nice theoretical
investigation is carried out in \cite{BDV}
and the one-loop contribution of Goldstone Bosons is also included.
The effective $W$ approximation is employed. The limits are based
on the observation at the $LHC/SSC$ of an excess of $50\%$ in the {\em total}
$W^+ W^+$ yield for an invariant $WW$ mass in the range $0.5< M_{WW} <1.$TeV.
Unfortunately the branching fraction into the 1st and 2nd generation leptons
are not included while this is essential to reconstruct these events!
Also
the irreducible \sm background is not taken into account. So as the authors
stress, the limits are subject to substantial uncertainties. More realistic
limits should be within an order of magnitude of those quoted in \cite{BDV}.

\subsection{Comparisons and Conclusions}
In Fig.~1 I show the limits one would obtain at the next colliders on the two
parameters $L_{9L}$ and $L_{9R}$ or equivalently using the conversion
in (\ref{conversion})
the parameters $L_W$ and $L_B$. These limits can also be interpreted,
in the case of the two-body reactions, as limits on the set
($\delta_Z\propto \Delta g_1^Z$,
$\Delta \kappa_\gamma$) \underline{with} the $SU(2)_c$
symmetry constraint $x_Z c_w=-x_\gamma s_w$ on $\Delta \kappa_Z$.
For those who prefer the latter parameterization,
the $L_{9L}$ axis is also the $\Delta g^Z_1 \propto \delta_Z$ axis, while
the $\Delta \kappa_\gamma$ axis is shown as $\Delta \kappa_\gamma=0$.
 ($\Delta \kappa_\gamma$ are the isolines $\Delta \kappa_\gamma\sim (L_{9L}
+ L_{9R})\times 1.35\;10^{-3}$).
Also, in
this respect, it is worth pointing out that the limits one gets on
$\Delta \kappa_{\gamma}$ when fitting
one parameter at a time, crucially depend on which
gauge-invariant operator, that is which model, $\Delta \kappa_{\gamma}$
originates from. The
discrepancy between limits on $\Delta \kappa_{\gamma}$ due to $L_{9L}$ and
$L_{9R}$ is even more drastic for the $pp$ machines. For instance
translating the limits on $\lnl$ ($\lnr$) as bounds on
$\dkg \equiv \dkg (\lnl)$ ($\dkg \equiv \dkg (\lnr)$), we have

\beqn
|\dkg (\lnl)| <3.\;10^{-3} \;\;&,&
-6.\; 10^{-3}< \dkg (\lnr) < 7.\; 10^{-3} \;\;\; \rm{(NLC(500))} \nonumber  \\
-2.\; 10^{-2} < \dkg (\lnl) < 10^{-2} \;\;&,&
-.17 < \dkg (\lnr) < 0.16 \;\;\;\;\;\; \rm{(SSC)}
\eeqn

\begin{table}[h]
\caption{\label{compareL9}{\em Expected limits on $L_{9L}$, $L_{9R}$ at LEP2,
the planned and next colliders.}}
\centering
\begin{tabular}{|l||c|c|c||c|c|c}
\cline{2-6}
\multicolumn{1}{c||}{} &LEP$190$GeV&NLC($500$GeV)&NLC($1$TeV)&SSC&LHC\\
\multicolumn{1}{c||}{}
&${\cal L}=500pb^{-1}$&${\cal L}=10fb^{-1}$&${\cal L}=44fb^{-1}$
&${\cal L}=10fb^{-1}$&${\cal L}=10fb^{-1}$\\
\multicolumn{1}{c||}{}
&\cite{BMT}&\cite{BMT}&\cite{BMT}&\cite{FLS}&\cite{FLS}\\
\cline{2-6}
\multicolumn{1}{c||}{}&\multicolumn{5}{c|}{}\\
\multicolumn{1}{c||}{}&\multicolumn{5}{c|}{{\bf \large 1-Parameter Fit}}\\
\hline
$L_{9L}$&$|L_{9L}|<30.4$&$|L_{9L}|<2.2$&$|L_{9L}|<0.7$&$-16\leftrightarrow7$&
$-22\leftrightarrow 12$\\ \hline
$L_{9R}$&$-125\leftrightarrow 155$&$-4.4\leftrightarrow 5.1$
&$-1.5\leftrightarrow 1.5$
&$-119\leftrightarrow 113$&$-152\leftrightarrow 147$\\ \hline
\multicolumn{1}{c||}{}&\multicolumn{5}{c|}{}\\
\multicolumn{1}{c||}{}&\multicolumn{5}{c|}{{\bf \large 2-Parameter Fit}}\\
\hline
$L_{9L}$&$-103\leftrightarrow 65$&$-13.4\leftrightarrow 7.7$&$-3.7< L_{9L}
<3.1$
&$-16\leftrightarrow7$&
$-22\leftrightarrow 12$\\
$L_{9R}$&$-260\leftrightarrow 760$&$-8.6\leftrightarrow 60.$&$-4.6
<L_{9R}<19.2$
&$-119\leftrightarrow 113$&$-152\leftrightarrow 147$\\ \hline
\hline
\end{tabular}
\end{table}

The $L_{9L}$ is always much better constrained than $L_{9R}$ especially in
$pp$ machines. In fact the limits one gets at $pp$ are almost an order
of magnitude worse for $L_{9R}$. As the figure shows (see also Table~3),
in the case of a
one-parameter fit, as if one were fitting to a particular model,
the $NLC500$ does much better than the SSC by more than a  factor $20$ on
$L_{9R}$ and $\sim 3 \div 5$ on $L_{9L}$. If two parameters are fitted, one
sees
that the best combined fit comes from the NLC where the $\gamma \gamma$ option
helps in reducing the range of the allowed
$L_{9L} \leftrightarrow L_{9R}$ parameter space
even further.  Compared to LEP2, NLC brings
an order of magnitude improvement, at least...\cite{BMT}.
The analysis also shows that
especially in the case of the two-prameter fit, the LEP2 limits translate into
large $L_{9}$ values ( $-260<L_{9R}< 760$). This is meaningless in
terms of a chiral expansion and may be that, after all, one should stick
with the phenomenological parameterization for such large values.
Although this presumes optimistic
expectations....\\
\noi A few words on the coupling $\la$. In the \epm environment
where the final polarizations can be reconstructed and where the $\la (L_\la)$
lead to essentially transverse states, these can be easily
disentangled from other
couplings. The limits on $\la$ from a one parameter fit or from a
three-parameter fit, as the {\em BM2} analysis shows, is not much different:
the limits are $\sim 10^{-2}$ at the NLC500. \\
\noi As the ``bars" in Fig. 1 show the $3V$ cross-sections do not
bring new constraints on the tri-linear couplings: the limits
are about an order of magnitude worse than in $WW$ production
at $500$GeV. Therefore the $3V$ reactions can be
``safely" exploited to look for the quartic couplings. \\
At the NLC500 we find the limits:
\beqn
-96.2 < \lu < 81.4 \;\;,\;\;&|\ld|&<118.4\;\;,\;\; 81.4<L_V<70.3 \nonumber \\
44<&L_{1,2}&<48
\eeqn
The last limit assumes that a $3Z$ final state has been identified taking a
$4\sigma$ deviation, while the first assumes a $3\sigma$ in $WWZ$
or the corresponding equal number of excess events in $ZZZ$.
Our preliminary study shows that with a
$1TeV \int \cl=60fb^{-1}$ \epm, these limits can be pushed to $\sim 6$.
They would
then compete with the SSC limits where the {\em theoretical} (see above)
analysis points to values of order $1$. \\

\noi In conclusion, a moderate energy \epm machine such as the NLC500
would bring an invaluable information on the symmetry breaking mechanism as
exemplified by the NLC bound in Fig.~1. Reconstruction of the various
W helicities is important and the \gag mode would be a very welcome addition.
In the case of models where the genuine
quartic couplings are smaller than
or of the same order as the tri-linear couplings (somehow ``vector dominated
models")
\footnote{These comparative discussions are with the tacit assumption
that operators which contribute to the 2-point function are not generated
by these models or that they are drastically suppressed! The optimistic
conclusions of this section would have to be ``watered down"
when this is not the case.
For instance, technicolour-like models naively mimicking QCD and including
heavy vector
resonances are of the vector-type. Unfortunately, they also predict a
contribution to the $2-W$
coupling at {\em tree-level} through the operator
$\cl_{10}=g g'\frac{L_{10}}{16 \pi^2} \tr(\Sigma B_{\mu \nu} \Sigma^\dagger
\W^{\mu \nu})$. This is related to the $S$ parameter\cite{Peskin}:
$L_{10} \ra -\pi S$. On the other the relations between the $L_i$ from
integrating $\rho$-like heavy vectors are
\cite{Gasser,FLS,BDV}: $L=L_{10}=-L_{9L}=-L_{9R}=4 L_1=-4 L_2$. The limit on
$L_{10}(M_Z)$ as extracted at the $M_Z$ scale from the Z data gives
$L_{10}(M_Z) \simeq -0.2 \pm 1.7$\cite{BarbieriChina}.
Following
\cite{BDV} and assuming that the previous relations between the $L_i$
hold at the scale $1.5TeV$ (mass of the vector) means a present
bound: $-1.4 <L_{10}<2.$ The
latter bound when compared to the limits on $L_{9L}$, $L_{9R}$ (see
Table~3) and $L_1$ means that these
models are already very much constrained by the LEP1 data and that the NLC500
would hardly improve on this limit. Ideally, one needs a 1TeV version of the
NLC (see Table~3).}
, the NLC500 seems to
be more constraining than the SSC. With only
$500GeV$, not allowing $WW$ scattering analyses, the NLC500 cannot compete
with the SSC in the case of the ``scalar models" which I associate
with models with
a ``preference" for $L_1, L_2$. From another viewpoint, the latter, in case
of a light Higgs or a ``not-too-heavy" Higgs, are expected to be
much much smaller
than the tri-linear couplings. This means that, even through these indirect
effects, the NLC500 is an excellent machine for a light or not-so-heavy
Higgs scenario. \\
Once again, by far, the best channel is $W$ pair production in \epm.
Theoretically this channel is also very ``clean" when compared to
the many uncertaintities
in the physics of $W$ at the $pp$ colliders. The full radiative corrections are
well under control \cite{RCWW} and good Monte-Carlo programs exist
\cite{Mannel}. The next step
which is now easy to implement, in order to get a more meaningful bound on
the parameters of the \nph, is to combine the \sm radiative corrections and
the ``anomalous" parameters together with the use
of the powerful fitting procedure of the
{\em BMT} Collaboration\cite{BMT}.

\vglue 0.3cm
{\bf \noindent Acknowledgements:}
I am most grateful to Misha Bilenky for providing me with the results of the
$L_{9L}-L_{9R}$ fit using the {\em BM2} programme and all the members of the
$W$ Working Group of the European \epm Workshop for discussions, suggestions
and
helpful criticisms. I also thank Genevi\`eve
B\'elanger and Marc Baillargeon for an enjoyable collaboration on $W$
physics.
\vglue 0.3cm

\end{document}